\documentclass[twocolumn,showpacs,preprintnumbers,amsmath,amssymb]{revtex4}
\usepackage{graphicx}
\usepackage{dcolumn}
\usepackage{bm}
\usepackage{epsfig}

\newcommand{\beq}{\begin{equation}}
\newcommand{\eeq}{\end{equation}}

\newcommand{\di}{\displaystyle}

\newcommand{\ga}{\gamma}

\newcommand{\scN}{{\scriptscriptstyle N}}
\newcommand{\scV}{{\scriptscriptstyle V}}

\newcommand{\QCD}{{\scriptscriptstyle QCD}}

\newcommand{\si}{\sigma}

\renewcommand{\epsilon}{\varepsilon}

\newcommand{\GeV}{\; \mbox{\rm GeV}}
\newcommand{\nb}{\; \mbox{\rm nb}}

\begin{document}

\title{Multiplicity of
photohadronization and photon--hadron scaling violation}

\author{Yury Novoseltsev}
\email{YuNovoselt@yandex.ru}
\author{Rita Novoseltseva}%
\affiliation{Institute for Nuclear Researsh of RAS,
 117312 Moscow, Russia}
\author{Grigory Vereshkov}
\email{gveresh@gmail.com}
\affiliation{
Research Institute of Physics, South Federal University,
 344090 Rostov-on-Don, Russia}
\affiliation{Institute for Nuclear Researsh of RAS,
 117312 Moscow, Russia}

\begin{abstract}
The method of scaling transformations permitting to carry out the
reconstruction of cross sections of $\ga N$ and $\ga \ga$
interactions on the basis of cross sections of
nucleon-(anti)nucleon interactions is suggested. The photon--hadron
scaling violation is a consequence of dependence of scaling
transformation parameter $\bar n(s)$ on the energy. The universal
function $\bar n(s)$ is interpreted as the multiplicity of
photohadronization. This function is established by processing the
data on $\ga p$ cross sections in the low energy region $\sqrt{s}<
20 \GeV$ and is extrapolated to the high energy region up to
$\sqrt{s}\sim 200\ \GeV$. The results of the reconstruction of
$\ga N$ cross sections at high energies and of $\ga \ga$ ones at
all energies are in a remarkable agreement with available
experimental data.

\end{abstract}

\maketitle

\section{Introduction}

In the present work we investigate the problem of the
photon--hadron scaling violation. Analysis of available
experimental data on cross sections of $\ga N$ interactions shows
that the scaling violation can be completely described within the
framework of the multiple photohadronization model. The model
development has led to the discovery of a new photon--hadron
symmetry which can be called {\it the local photon--hadron
scaling}. The question is the relation between cross sections of
$\ga N$ and $\ga \ga$ interactions and the ones of hadron--hadron
interaction:
\beq
\begin{array}{c}
\displaystyle \si_{\ga \scN}(s)=\bar n(s)\cdot \si_{\ga \scN}^{(0)}(s/\bar n(s)),
\\[3mm] \displaystyle \si_{\ga \ga}(s)=\bar n^2(s)\cdot \si_{\ga \ga}^{(0)}(s/\bar n^2(s)),
\end{array}
\label{1}
\eeq
where $s$ \ is the energy squared in the
center--of--mass system, $\si_{\ga \scN}(s)$ and $\si_{\ga \ga}(s)$
are cross sections of $\ga N$ and $\ga \ga$ interactions, $\si_{\ga
\scN}^{(0)}$ and $\si_{\ga \ga}^{(0)}$ are so--called calibration
functions whose form is determined on the basis of processing the
data on cross sections of nucleon--(anti)nucleon interactions. The
parameter $\bar n(s)$ is the {\it universal} function describing the
photohadronization multiplicity.
This function and available hadron data allow to
reconstruct the total cross sections of $\ga N$ and $\ga \ga$
interactions up to energies
$\sqrt{s_{\ga \scN}},\ \sqrt{s_{\ga \ga}} \sim $ 1000 GeV via the scaling transformaion (\ref{1}).

In Section 2 we bring forward a more accurate formulation of
scaling relations and construct the calibration function $\si_{\ga
\scN}^{(0)}$. In Section 3 we propose the model of multiple
photohadronization.
The analytical form of the photohadronization multiplicity, $\bar n(s)$, is imported from
QCD calculations. Parameters of this function together with
a criterion of its universality are established in Section 4. In
Section 5 the same function is used for the reconstruction of
cross section of $\ga \ga$ interaction.

\section{Experimental data}\label{exp}

Total cross sections of $\ga p$ interaction
are measured in a wide energy region from $\sqrt{s}\sim 1\ \GeV$
to $\sqrt{s}=210\ \GeV$. There are direct accelerator data of high
precision and completeness at low energies. In what follows, we
shall use the data obtained in \cite{1} at $\sqrt{s}= 5.93\, -\,
18.54\ \GeV$. In this region the typical value of the cross
section is $\sigma_{\ga p}\approx 115\ \mu b$. At $\sqrt{s}\sim
200\ \GeV$ there are results of four measurements obtained in
DESY:
  \[
 \begin{array}{c}
\displaystyle \sqrt{s},\ \GeV \quad\quad 180{\phantom{\pm 11}} \quad 210{\phantom{\pm 11}} \quad  200{\phantom{\pm 11}} \qquad 210{\phantom{\pm 11}}
 \\[3mm]
\displaystyle \sigma_{\ga p},\ \mu b\quad\quad 143\pm 4 \quad 154\pm 16 \quad 165\pm 2.3
 \quad 174\pm 1
 \end{array}
 \]
 The measurement at $\sqrt{s}=180\ \GeV$ and two ones at
$\sqrt{s}=210\ \GeV$ have been carried out by ZEUS collaboration
\cite{2,3}. The cross section at $\sqrt{s}=200\ \GeV$ was measured
by H1 collaboration \cite{4}. Database \cite{5} includes H1 result
and one of ZEUS ones: $\sigma_{\ga p}= 174\pm 1\ \mu b$ at
$\sqrt{s}=210\ \GeV$ (two other results are likely to be
understated). Further, database \cite{5} includes cross sections
measured by Baksan collaboration \cite{6} in intermediate energy
region:
 \[
 \begin{array}{c}
 \displaystyle \sqrt{s},\ \GeV{\phantom{111}}\; 45{\phantom{\pm 11}} \qquad 54{\phantom{\pm11}}\qquad
   75{\phantom{\pm11}} \qquad
 134{\phantom{\pm11}}
 \\[3mm]
 \displaystyle \si_{\ga \scN},\ \mu b\quad 128\pm 14 \quad 146\pm 15 \quad
  148\pm 23
 \quad 152\pm 80
 \end{array}
 \]
Experimental data on cross sections of $\ga p$ interaction are
presented in Fig.1. Fig.2 displays data on cross sections of
$\ga \ga$ interaction that we shall discuss in Section 5. As it is
evident from the data shown in Fig.1, there is a rapid rise of
$\ga p$ cross section with the rise of energy. This fact is known
as the effect of the photon--hadron scaling violation. The problems of photon--hadron scaling
 and its violation were discussed, for example, in ~\cite{7,8,9}. It is
customary to refer to the relation below, which follows from
additive quark model, as "scaling":
 \beq
 \di
 \sigma_{\gamma p}^{(0)}(s)=P_{\gamma\to \scV}\times \frac23\bar
 \sigma_{\scN p}(3s/2),
 \label{2}
 \eeq
where $P_{\ga \to \scV}=1/250$ is the probability of photon
hadronization to vector mesons $\rho(770),\ \omega(782),\
\phi(1019)$, and
 \beq
 \displaystyle
 \bar\sigma_{\scN p}=\frac14\left(\sigma_{pp}+\sigma_{\bar pp}+\sigma_{np}+\sigma_{\bar
 np}\right).
 \label{3}
 \eeq
Factors of $3/2$ appear in (\ref{2}) due to recalculation of the
interaction cross section of a 3--quark system with a 3--quark one
to the interaction cross section of a 2--quark system with a
3--quark one \cite{6}. The quark diagram representing $\ga
N$ interaction in the photon--hadron scaling mode is shown in Fig.3a.

We shall call the function $\sigma_{\gamma p}^{(0)}(s)$ (\ref{2})
the calibration curve for the cross section of $\ga
p$ interaction. Clearly, the deviation of measured cross sections
$\si_{\ga p}(s)$ from the calibration curve is a quantitative
characterization of the scaling violation.

For constructing the calibration curve in the energy region
$\sqrt{s}=5.93\; -\; 22.97$ GeV, we used data from \cite{10,11}.  In the
region $\sqrt{s}=30.4\; -\; 62.7$ GeV, where
$\sigma_{pp}=\sigma_{np}$\ , $\sigma_{\bar pp}=\sigma_{\bar pn}$
with good accuracy, we used only the data on $\sigma_{pp}(s)$ and
$\sigma_{\bar pp}(s)$ from \cite{12,13}. For collider energies
$\sqrt{s}=200,\ 546,\ 900,\ 1800\ $ GeV, the $\bar p p$ data from
\cite{14,15,16,17} were used. Experimental data, used to construct
the calibration curve (after scaling transformation (\ref{2})),
are presented in Fig.1. In the same figure, the fit of the data
on averaged cross sections (\ref{3}), subjected to the scale
transformation $3s_{\scN p}/2=s_{\ga p}\equiv s$, is shown. The
fitting function and values of fitting parameters are
 \beq
 \begin{array}{c}
 \displaystyle \sigma_{\ga p}^{(0)}(s)=
 C_{\ga p}\ln^2\frac{s}{s_0}+A_{\ga p}\left(1+\frac{s_{\ga p}^{(1)}}
 {s+s_{\ga p}^{(2)}}\right),
 \\[5mm]\displaystyle
 C_{\gamma p}=0.5777\ \mu b,\qquad \sqrt{s_0}=2.198\ \GeV,
 \\[3mm] \di A_{\gamma p}=95.65\ \mu
 b,
 \\[3mm]
 \displaystyle
 \sqrt{s_{\ga p}^{(1)}}=
 2.774\ \GeV,\qquad \sqrt{s_{\ga p}^{(2)}}=3.589\ \GeV.
 \end{array}
 \label{4}
 \eeq
It should be stressed that $\sigma_{\gamma p}^{(0)}(s)$ has the
status of an {\it experimentally} established function. Its
further usage in the analytical form (\ref{4}) is motivated, first
of all, by high quality of the fit. Additional physical grounds
related to, for example, Froissart asymptotics  have a certain
sense, but they do not have crucial meaning for the problem under
discussion.
\begin{figure}
    \begin{minipage}[b]{1.0\linewidth}
        \centering\epsfig{figure=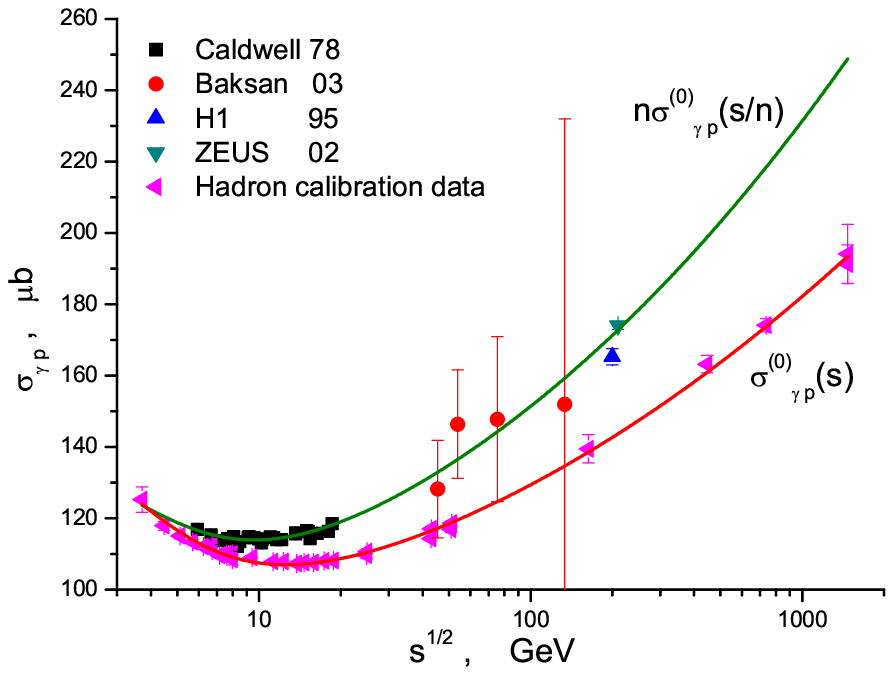,width=\linewidth}
        \caption{Total cross section of $\ga N$ interaction.
  $\si_{\ga p}^{(0)}(s)$ is the calibration curve (\ref{4}).
  The upper line is the cross section $\si_{\ga \scN}(s)$ according
  to (\ref{6}).}
        \label{ris1}
        \end{minipage}
\end{figure}

\begin{figure}
    \begin{minipage}[b]{1.0\linewidth}
        \centering\epsfig{figure=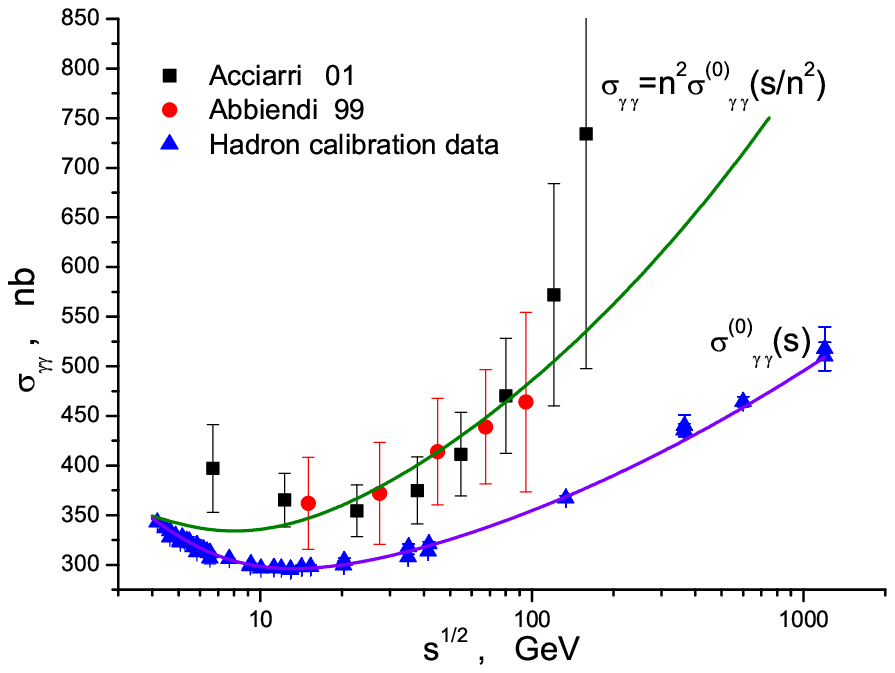,width=\linewidth}
        \caption{Total cross section of $\ga \ga$ interaction.
       $\si_{\ga \ga}^{(0)}(s)$ is the calibration curve (\ref{10}).
       The upper line is the cross section $\si_{\ga \ga}(s)$ according
  to (\ref{11}).}
        \label{ris2}
        \end{minipage}
\end{figure}

\begin{figure}
    \begin{minipage}[b]{.96\linewidth}
        \centering\epsfig{figure=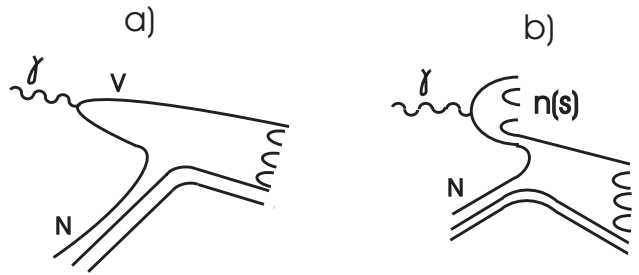,width=\linewidth}
\caption{Quark diagrams of $\ga N$ interaction: a) photon-hadron
scaling mode, b) multiple photohadronization model.}
        \label{ris3}
        \end{minipage}
\end{figure}

\section{Scaling violation model}\label{sc-v}

Photon--hadron scaling violation has a simple interpretation and,
as we will show below, it is described by the universal function
whose form is predicted by Quantum Chromodynamics (QCD). The main
idea of the suggested model is illustrated in Fig.3b. We assume
that the photon hadronization in the nonperturbative vacuum

(i) firstly, takes place before the interaction of hadronization
products with a nucleon,

(ii) secondly, as a result of
hadronization, a hadron cluster (mainly mesons) with total quantum
numbers $J^{PC}=1^{--}$ \ appears.

As one can see from Fig.3b,
the multiplicity of $\ga N$ interaction is produced with both the
multiplicity of photohadronization process and the multiplicity of
particles produced in the interaction of one of the mesons of the
cluster with the nucleon. The latter means that the multiplicity
of the photohadronization process is described by a function which
has the same mathematical structure as the multiplicity function
of processes of hadron--hadron interactions and $e^+e^-$
annihilation to hadrons. There are several different
representations of the multiplicity function which are consistent
with the experimental data. QCD calculations result in the
functions of the following form:
 \beq
 \begin{array}{c}
 \displaystyle \bar n(s)=a+b\exp\left[c\sqrt{\ln(s/\Lambda^2)}\right],
 \\[3mm]
 \displaystyle \bar n(s)=b\ln^a(s/\Lambda^2)\exp\left[c\sqrt{\ln(s/\Lambda^2)}\right],
 \end{array}
 \label{5}
 \eeq
where $\Lambda$ is a cutoff parameter whose value is determined by
the hadronization threshold; $a,\ b$ and $c$ are fitted
parameters. The first of the functions (\ref{5}) was suggested in
\cite{18}, the second one -- in \cite{19}. Note the functions
(\ref{5}) coincide with each other at $a=0$. Using these functions
to describe the multiplicity of photon hadronization we discovered
that in the experimental data fitting the value of the parameter
$a$ automatically turns out to be small and the value of cutoff
parameter $\Lambda$ is automatically close to
$\Lambda_{\QCD}\simeq 2m_\pi$. The results of the four--parameter
fit turn out to be statistically equivalent to the results of the
two--parameter fit over parameters $b$ and $c$ with fixed values of
$a=0$, $\Lambda=2m_\pi$. Because of this, we suggest the following
formula for the cross section of $\ga N$ interaction for the model
presented in Fig.3b:
 \beq
 \begin{array}{c}
 \displaystyle \si_{\ga \scN}(s)=\bar n(s)\cdot \si_{\ga
 p}^{(0)}(s/\bar n(s)),
 \\[3mm]
 \displaystyle \bar n(s)=
 b\exp\left[c\sqrt{\ln(s/4m_\pi^2)}\right].
 \end{array}
 \label{6}
 \eeq

The function $\si_{\ga p}^{(0)}(s_n)$ in (\ref{6}) depends on its
argument $s_n\equiv s/\bar n(s)$ in the same way as the function
(\ref{4}) depends on the variable $s$. The scaling transformation
$s\to s_n$ takes into account a redistribution of the energy of
the primary photon between products of its hadronization. The
values of parameters $b$ and $c$ are established by comparing the
model with the experimental data.

Note that (\ref{6}) belongs to the additive quark model. The
starting point is the expression for the cross section of an
interaction of a nucleon, considered as a 3--quark system, with a
2--quark system. This expression should be subjected to the scaling
transformation to the interaction cross section of a nucleon with
a $2\bar n$-quark system. The violation of photon--hadron scaling
appears because the average number of quarks, which are produced
in a photon hadronization and interact with a nucleon, increases
with the energy.

\section{Data fit. A criterion of universality of the mechanism of
photon-hadron scaling violation}\label{fit}

We carried out two fitting procedures. In the first one we took
into account all data on cross sections of $\ga p$ interaction
over the whole energy range plotted in Fig.1. The following
values for the fitted parameters have been obtained:
 \beq
 \begin{array}{c}
 \displaystyle b=0.713\pm 0.011,\qquad c=0.148\pm 0.006,
 \\[3mm]\displaystyle
 \chi^2/dof=0.949.
 \end{array}
 \label{7}
 \eeq
In second procedure we used only the low-energy data obtained in
\cite{1} in the region $\sqrt{s}= 5.93\, -\, 18.54\ \GeV$. The
results are
 \beq
 \begin{array}{c}
 \displaystyle b=0.714\pm 0.025,\qquad c=0.147\pm 0.013,
 \\[3mm]\displaystyle
 \chi^2/dof=0.819.
 \end{array}
 \label{8}
 \eeq
As is seen from (\ref{7}), (\ref{8}), the values of fitted
parameters obtained in different procedures are practically the
same. When we extrapolate the fitting curve built on "low--energy"
parameters (\ref{8}) into the region of intermediate and high
energies, it almost coincides with the curve whose parameters
(\ref{7}) were obtained by fitting all the data.

The possibility of the prediction of quantitative characteristics
of photon--hadron scaling violation is a distinguishing and quite
natural feature of the suggested model. Indeed, at the parton
level, patterns of statistic process of pfotohadronization {\it
have to be} completely {\it formed} at comparatively low energies
--- only slightly greater than the production threshold of strange
hadrons. Parton cascades evolution {\it has} to be the universal
function of the interaction energy whose analytical form is the
same in the whole energy range.

The agreement of results of fits (\ref{7}) and (\ref{8}) can be
interpreted as an argument supporting the hypothesis of multiple
photohadronization and in the same time as a confirmation of the
possibility to describe this phenomenon via the universal function
of an energy of $\ga N$ interaction.

\section{Reconstruction of a cross section of $\ga\ga$ interaction}\label{ga-ga}

Dependence of a cross section of $\ga\ga$ interaction in the
photon--hadron scaling approximation is described by the following
formula
 \beq
 \displaystyle \si_{\ga \ga}^{(0)}(s)=P_{\ga \to
 \scV}^2\times\frac49\si_{\bar pp}(9s/4).
 \label{9}
 \eeq
We use in (\ref{9}) only cross sections of $\bar pp$
interactions, taking into account the similarity of
quark--antiquark structures of $\bar pp$ and $\ga\ga$ pairs.
Factors $9/4$ appear in (\ref{9}) as the result of recalculation
of the interaction cross section of 3--quark systems $\bar pp$ to
the one of 2--quark systems $VV$.

For constructing the calibration curve $\si_{\ga \ga}^{(0)}(s)$,
we used all experimental data on $\si_{\bar pp}(s)$ at $\sqrt{s}
\ge 2.79\ \GeV$. After applying scaling transformation (\ref{9})
to the data, we fitted them using the formula with Froissart
asymptotics. The results are:
 \beq
 \begin{array}{c}
 \displaystyle \si_{\ga \ga}^{(0)}(s)=
 C_{\ga \ga}\ln^2\frac{s}{s_0}+
 A_{\ga \ga}\left(1+\frac{s_{\ga \ga}^{(1)}}{s+s_{\ga \ga}^{(2)}}\right),
 \\[5mm]\displaystyle
 C_{\ga \ga}=1.541\ \nb,\qquad \sqrt{s_0}=2.198\ \GeV,
 \\[3mm] \di A_{\ga \ga}= 264.525\
 \nb,
 \\[3mm]
 \displaystyle
 \sqrt{s_{\ga \ga}^{(1)}}=2.855\ \GeV,\qquad \sqrt{s_{\ga \ga}^{(2)}}=3.385\ \GeV.
 \end{array}
 \label{10}
 \eeq
Asymptotically at $s\to \infty$, cross sections in all channels of
nucleon--(anti)nucleon interactions are equal. Therefore the
following relation holds:
\[
\displaystyle C_{\ga \ga}=\frac23P_{\ga \to \scV}C_{\ga p}
\]
Thereby the parameter $C_{\ga \ga}$ in (\ref{10}) is fixed by the
results of the fit (\ref{4}). The parameter $s_0$ is taken to be
common in (\ref{4}) and (\ref{10}) because this assumption does
not contradict the experimental data. In fact, only three
parameters $A_{\ga \ga},\ s_{\ga \ga}^{(1)},\ s_{\ga \ga}^{(2)}$
were fitted.

The experimental data on cross sections of $\ga \ga$ interaction
\cite{20,21} are presented in Fig.2. The scaling transformated
experimental data, that were used for constructing the calibration
curve (\ref{9}), are shown in the same figure. The fit results of
the data from the formula (\ref{10}) show that this formula has
the status of an experimentally established function.

Within the framework of the hypothesis of multiple
photohadronization the formula for a cross section of $\ga \ga$
interaction is built by an obvious scaling transformation
 \beq
 \begin{array}{c}
 \displaystyle \si_{\ga \ga}(s)=\bar n^2(s)\cdot \si_{\ga
 \ga}^{(0)}(s/\bar n^2(s)),
 \\[3mm]
 \displaystyle \bar n(s)=
 b\exp\left[c\sqrt{\ln(s/4m_\pi^2)}\right].
 \end{array}
 \label{11}
 \eeq
The function $\sigma_{\ga \ga}^{(0)}(s_n)$ in (\ref{11}) depends
on its argument $s_n\equiv s/\bar n^2(s)$ in the same way as the
function (\ref{10}) depends on the variable $s$. In (\ref{11}) and
(\ref{6}) $\bar n(s)$ is the same function with the same values of
parameters $b$ and $c$. In such a way, one can reconstruct cross
sections of $\ga \ga$ interaction using the processed experimental
data on cross sections of $\ga N$ interaction (see the
corresponding curve in Fig. 2). The reconstruction results agree
with the experimental data at the level of $\chi^2\sim 0.5$. Only
cross section values for $\sqrt{s}$ = 120.4 and 158.7 GeV
noticeably deviate from the prediction of formula (\ref{12}).
However, these cross sections are measured with large systematic
errors.

\section{Discussion and conclusions}\label{dis}

Cross sections of $\ga N$ and $\ga \ga$ interactions can be
obtained from hadron cross sections via scaling transformations
(\ref{2}), (\ref{6}) and (\ref{9}), (\ref{11}). The photon--hadron
scaling violation is completely described by dependence of the
scaling transformation parameter $\bar n(s)$ appearing in
(\ref{6}), (\ref{11}) on the energy. The universal function $\bar
n(s)$ is interpreted as the multiplicity of a photohadronization.
Parameters of this function are determined by processing the
experimental data on cross sections of $\ga p$ interaction in the
region of low energies. Once the function $\bar n(s)$ is known,
cross sections of $\ga N$ interaction at high energies and the
ones of $\ga \ga$ interaction at all energies can be reconstructed
using the suggested model in complete agreement with available
experimental data.

The results obtained in the present work show that the
photon--hadron scaling problem can be formulated in new terms.
Formulas (\ref{2}) and (\ref{9}) express the idea of the {\it
global} scaling and numerical parameters in these formulas ($2/3$
and $4/9$ respectively) make sense of parameters of {\it global}
scaling transformations. Formulas (\ref{6}) and (\ref{11}), which
describe experimental data, allow us to introduce a peculiar
photon--hadron symmetry, namely, {\it the local photon--hadron
scaling} with scaling transformation parameters $2\bar n(s)/3$ and
$4\bar n^2(s)/9$ in $\ga N$ and $\ga \ga$ channels respectively.
It is important that the dependence of {\it local} scaling
transformation parameters on the interaction energy is determined
by the universal function $\bar n(s)$. The idea of local
photon--hadron scaling is expressed explicitly by formulas
connecting photon and hadron cross sections (we use (\ref{2}),
(\ref{6}) and (\ref{9}), (\ref{11}))
 \beq
\begin{array}{c}
\displaystyle
 \si_{\ga \scN}(s)=P_{\ga\to \scV}\times \frac{2\bar n(s)}{3}\bar
 \si_{Np}\left(\frac{3s}{2\bar n(s)}\right),
 \\[5mm]
\displaystyle \si_{\ga \ga}(s)=P_{\ga \to
 \scV}^2\times\frac{4\bar n^2(s)}{9}\sigma_{\bar pp}\left(\frac{9s}{4n^2(s)}\right)\ .
 \end{array}
\label{12}
 \eeq
As one can see from Figures 1 and 2, the experimental data support
the model of local photon--hadron scaling up to energies
$\sqrt{s_{\ga \scN}} \approx $ 200 GeV and $\sqrt{s_{\ga \ga}}
\approx $ 100 GeV. Formulas (\ref{12}) and available hadron data
allow one to predict the photon cross sections up to energies
$\sqrt{s_{\ga \scN}},\ \sqrt{s_{\ga \ga}} \sim $ 1000 GeV. The
validity of local photon--hadron scaling in the region of LHC and
NLC energies is a prerogative of future experiments.

\end{document}